\documentclass[review]{elsarticle}

\usepackage{lineno,hyperref}
\modulolinenumbers[5]
\usepackage{amsmath,amsfonts,amssymb}
\hypersetup{colorlinks=true,linkcolor=blue,urlcolor=blue,citecolor=blue}

\journal{Icarus}

\usepackage{numcompress}\biboptions{authoryear}

\usepackage{ulem}

\begin{document}

\begin{frontmatter}

\title{Dynamical environments of MU69 and similar objects}


\author[add1]{Guillaume Rollin\corref{corr}}
\cortext[corr]{Corresponding author}
\ead{guillaume.rollin@utinam.cnrs.fr}

\author[add2]{Ivan~I.~Shevchenko}
\ead{i.shevchenko@spbu.ru}

\author[add1]{Jos\'e Lages}
\ead{jose.lages@utinam.cnrs.fr}

\address[add1]{Institut UTINAM, Observatoire des Sciences de l'Univers THETA, CNRS, Universit\'e
Bourgogne Franche-Comt\'e, Besan\c{c}on 25030, France}
\address[add2]{Saint Petersburg State University, 7/9 Universitetskaya nab., 199034 Saint
Petersburg, Russia}

\begin{abstract}
We explore properties of the long-term dynamics of particles
(moonlets, fragments, debris etc.) around KBO 2014 MU69 
(Arrokoth), as well as around similar contact-binary objects
potentially present in the Kuiper belt. The chaotic diffusion of
particles inside the Hill sphere of MU69 (or, generally, a similar
object) is studied by means of construction of appropriate
stability diagrams and by application of analytical approaches
generally based on the Kepler map theory. The formation and
evolution of the particle clouds, due to the chaotic diffusion
inside the Hill sphere, are studied and the cloud lifetimes are
estimated.
\end{abstract}


\end{frontmatter}


\section{Introduction}

The second (after Pluto) target object for the New Horizons
mission was chosen in 2014, finalizing an observation survey
performed with the Hubble Space Telescope \citep{S17}. It was
called 2014~MU69, and, subsequently, (486958) Arrokoth 
(temporarily it was also nicknamed Ultima Thule). Later on, due
to results of dedicated observational campaigns \citep{S17,P17},
this object was suspected to be a classical KBO, a primordial
contact binary (hereafter CB). A dumbbell contact-binary shape is
typical for KBOs. However, up to the time of the New Horizons
flyby, its lightcurve monitoring had not succeeded to retrieve the
rotation period, because visual magnitude variations were
unresolved \citep{B16}.

The rendezvous of New Horizons and Arrokoth took place on January
1, 2019. Indeed, Arrokoth turned out to be a contact binary
\citep{S19LPI,C19LPI,P19LPI,Stern19Sci}, visually fitting a
dumbbell model, depicted, e.g., in Fig.~5 in \cite{Sch07I}, or
most similar, in Fig.~1 in \cite{LSS17}.\footnote{Later on, the
Arrokoth constituents were reported to be flattened
\citep{Stern19Sci}.} Tantalizingly, the ratio of masses of the
binary components has turned out to be $\sim$1/3, very similar to
the ratios typical for contact-binary cometary nuclei, as compiled
in Table~1 in \cite{LSR18}.

Identifying any material in the vicinities of a target object of a
space mission is of an especial concern for planning cosmic
flybys, including that by Arrokoth \citep{M18GRL}, as the material
is hazardous for a space probe. Low-mass shallow matter orbiting
around Arrokoth, as around any other KBO, may originate from a
number of processes. It may be left from a primordial swarm of
solids \citep{MK19LPI}, or it may be ejecta of various origins:
ejecta due to early out-gassing \citep{T15AA,SL00JGR}; ejecta from
impacting by intruding bodies \citep{N18AJ}; ejecta resulting from
the CB-forming collision \citep{U19LPI}. However, up to now, no
moons, moonlets, fragments, or debris
\citep{K18,SSL19LPI,SSM20Sci}, or any traces of coma
\citep{G19LPI}, have been discerned in not-yet-completed image
surveys, performed from HST and New Horizons in the field around
Arrokoth.

As shown in \cite{LS20}, based on preliminary data on the shape of
Arrokoth, this rotating CB is able to efficiently cleanse its
vicinities by chaotizing all material orbiting it sufficiently
close. In this article, we explore properties of the long-term
dynamics of low-mass matter (whatever it can be: moonlets,
fragments etc.) around CB-shaped objects, expected to be
ubiquitous in the Kuiper belt.

To assess a global picture of the dynamical environment of
Arrokoth or a similar object, it is necessary (1)~to analyze the
process of cleansing of the circum-binary chaotic zone; (2)~to
analyze the process of formation and further survival of a cocoon,
formed by the ejected matter inside CB's Hill sphere. In this
article, our study is concentrated on just these two items.
Therefore, we are interested in the timescale of clearing the
immediate vicinity of Arrokoth (the chaotic circum-binary zone),
the possibility and timescale of formation of a cocoon of ejected
matter around Arrokoth inside its Hill sphere, and the
survivability of such a cocoon.

In our study, we aim to assess the rate of clearing process
in the chaotic circumbinary zone; to obtain the mass parameter
dependence of the depopulation rate; to estimate the
characteristic time of dispersal of low-mass matter out from
Arrokoth's Hill sphere, if such matter were initially present; to
assess collisional hazards for space probes visiting neighborhoods
of Arrokoth-like objects in the Kuiper belt.

\section{Circum-CB clearing: the problem setting}

Spinning gravitating CB-shaped bodies create zones of dynamical
chaos around them \citep{LSS17}, and this has a clearing effect:
any material put in orbits around a rotating dumbbell (e.g., any
material ejected from its surface) cannot survive in this chaotic
zone. It either escapes into space, or is absorbed by the parent
body's surface \citep{LSR18}. As the orbiting matter is removed in
this way, a spinning gravitating CB cleans-up its vicinities.

A much more well-known example of analogous ``cleansing'' is the
formation of the gap in the close-to-coorbital neighbourhood of a
planet \citep{W80,DQT89,MM15}. The close-to-coorbital chaotic gap
is formed by the overlap of the first-order mean-motion resonances
accumulating in the neighbourhood of a planet's orbit; whereas the
circum-CB chaotic zone is formed by the overlap of the
accumulating integer spin-orbit resonances with the rotating
dumbbell \citep{LSR18}. In the both cases, any material injected
into the chaotic zones is subject to an unlimited chaotic
diffusion in the eccentricity (as well as subject to possible
close encounters with the CB or the planet) and, therefore,
finally is removed.

\begin{figure}[!t]
\centering
\includegraphics[width=0.55\textwidth]{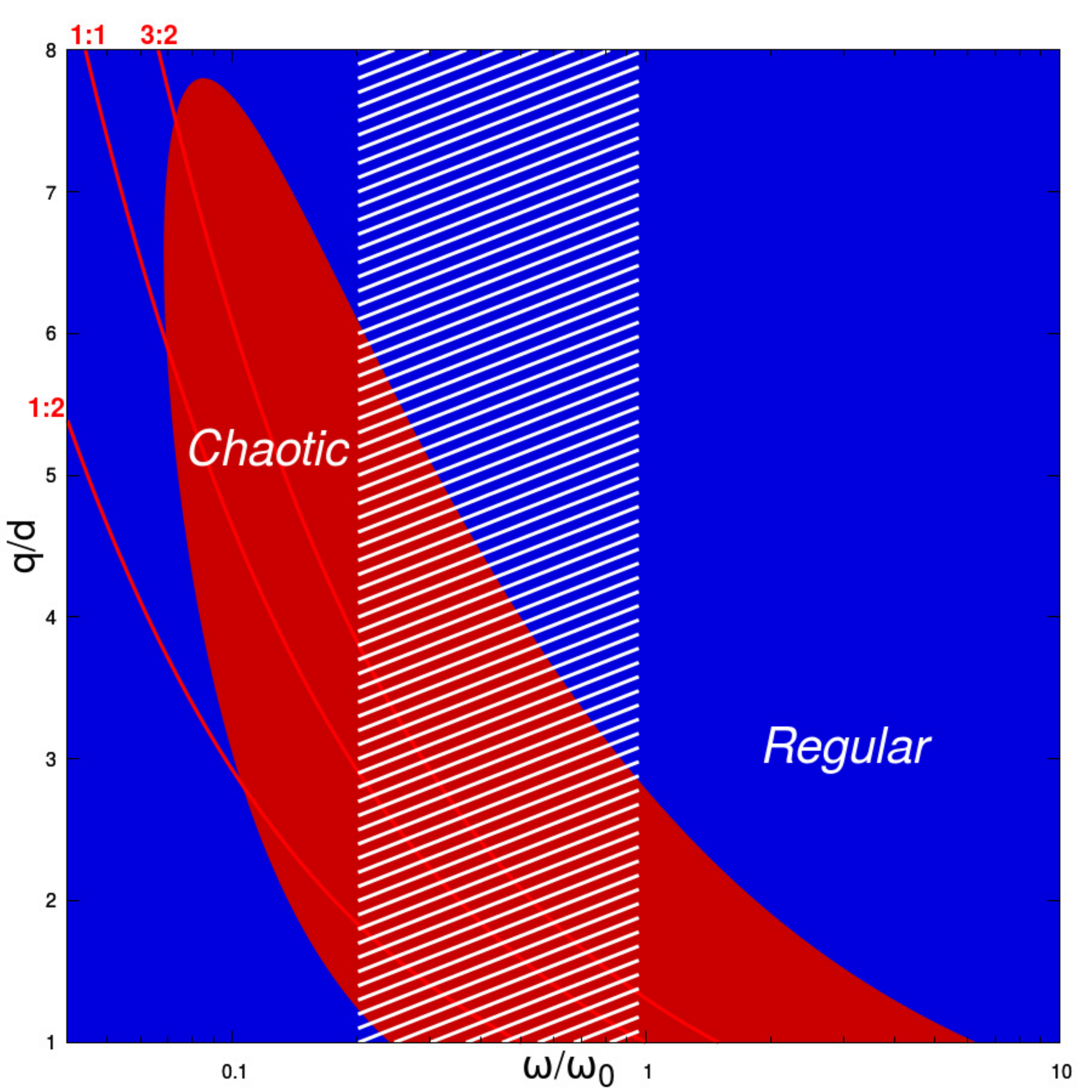}
\caption{Extents of the chaotic zone (shown in red) around a contact binary as a function of the
binary's rotation rate $\omega$, in ratio to the critical $\omega_0$. The pericentric distance $q$
is measured in units of $d$, the contact binary size. White shaded area delimits the range of
typical rotation rates of the Kuiper belt objects, according to data in \cite{TNO14}.
Red solid curves show locations of three major spin-orbit resonances.}
\label{Fig1}
\end{figure}

In Fig.~\ref{Fig1}, adapted from our previous study \citep{LSR18}, we represent graphically the
extents of the circum-CB chaotic zone. The diagram is set in the ``CB rotation rate -- particle's
initial pericentric distance'' frames. The rotation rate $\omega$ is measured in units of its
critical value $\omega_0$, corresponding to centrifugal disintegration of the initially-contact
binary ($\omega_0$ is equal to CB's Keplerian rate of rotation). The pericentric distance $q$ is in
units of the binary's size $d$, defined as the distance between the mass centers of its components.
In units of the critical rate $\omega_0$, the typical rotation rates $\omega$ of the Kuiper belt
objects range from 0.2 to 1 (thus, the periods range from 1 to 5, in critical periods), according
to the observational (lightcurve) data given in \cite{TNO14}. The area bounded by these limits in
Fig.~\ref{Fig1} is white-shaded. Locations of main resonances 1:2, 1:1, and 3:2 between orbiting
particles and the rotating central body are shown as red curves. Fig.~\ref{Fig1} demonstrates
that typical Kuiper belt CBs may have rather extended circum-body chaotic zones: for orbits inside
such zones, the initial pericentric distance $q$ ranges up to $\sim 6d$.

Recall that the radius of a gravitating body's Hill sphere
$R_\mathrm{H}$, in units of the semimajor axis of a perturber,
$a_0$, is given by

\begin{equation}
\frac{R_\mathrm{H}}{a_0} = \left( \frac{m}{3 M} \right)^{1/3} ,
\label{Hill_sma}
\end{equation}

\noindent where $M$ and $m$ are the primary's and secondary's
masses, respectively (those of the Sun and Arrokoth, in our
problem). The orbit of Arrokoth's any moonlet should lie within
Arrokoth's Hill sphere. This implies the inequality $a(1 + e)
\lesssim R_\mathrm{H}$.

Given the ``dumbbell size'' of Arrokoth $d \simeq 17$~km
\citep{Stern19Sci,McKi20}, it is straightforward to
estimate, using the diagram, that the chaotic clearing zone
around Arrokoth may have radius of at most $\sim$100 km, an order
of magnitude less than the New Horizons flyby distance ($\sim$3500
km) and three orders of magnitude less than Arrokoth's Hill radius
($\sim 5 \cdot 10^4$~km).

\section{Numerical simulations: the stability diagram}
\label{sec_stab}

To describe the immediate dynamical environments of Arrokoth, we
construct stability charts in the $q$--$e$ (pericentric distance
-- eccentricity) plane of initial conditions. We use the Lyapunov
characteristic exponent (LCE) method, that we have earlier
employed in \cite{LSS17,LSR18}. We choose an inertial Cartesian
coordinate system with the origin at the CB's center of mass. The
equations of motion of the particle with coordinates $(x, y)$ are
given by

\begin{equation}
\begin{array}{ccl}
\dot{x}&=&v_x, \\
\dot{y}&=&v_y, \\
\dot{v}_x&=&-{{{\it m_2}\,\left(x-{\it x_2}\right)}\over{\left(\left(y-
{\it y_2}\right)^2+\left(x-{\it x_2}\right)^2\right)^{{{3}/{2}}}
}}-{{{\it m_1}\,\left(x-{\it x_1}\right)}\over{\left(\left(y-
{\it y_1}\right)^2+\left(x-{\it x_1}\right)^2\right)^{{{3}/{2}}}
}},\\
\dot{v}_y&=&-{{{\it m_2}\,\left(y-{\it y_2}\right)}\over{\left(\left(y-
{\it y_2}\right)^2+\left(x-{\it x_2}\right)^2\right)^{{{3}/{2}}}
}}-{{{\it m_1}\,\left(y-{\it y_1}\right)}\over{\left(\left(y-
{\it y_1}\right)^2+\left(x-{\it x_1}\right)^2\right)^{{{3}/{2}}}
}},
\end{array}
\label{motion1}
\end{equation}

\noindent where $(x_1, y_1)$ and $(x_2, y_2)$ are the coordinates of the centers of masses $m_1$
and $m_2$, respectively. The locations $x_1, y_1$ and $x_2, y_2$ of the primaries are given by

\begin{equation}
\begin{array}{ccl}
x_1&=&\mu\cos(\omega t), \\
y_1&=&\mu\sin(\omega t), \\
x_2&=&(\mu-1)\cos(\omega t),\\
y_2&=&(\mu-1)\sin(\omega t).
\end{array}
\label{motion2}
\end{equation}

\noindent The quantity $\omega$ is a parameter responsible for the arbitrary rotation
frequency of
the CB; $\omega$ is equal to CB's rotation rate in units of
its critical rotation rate corresponding to
centrifugal disintegration. At $\omega = 1$, the equations reduce to the usual equations of motion
in the planar restricted three-body problem. The distance between the centers of masses
$m_1$ and $m_2$ is set here to unity, $d=1$. Also we set $\mathcal{G}(m_1+m_2)=1$; therefore, the
angular rate of the Keplerian orbital motion of the binary (if it were unbound) is

$$ \omega_0 = \left( \mathcal{G}(m_1+m_2)/d^3 \right)^{1/2} = 1 . $$

\noindent Arrokoth constitutes an alliance of two round
bodies\footnote{Although flattened; but as soon as the Arrokoth
components are flattened mostly orthogonal to its rotation plane
\citep{McKi20}, this flattening does not compromise our dumbbell
model for the gravitational potential.}; therefore, the dynamical
model given by Equations~(\ref{motion1})--(\ref{motion2}) is
expected to be essentially adequate.

We set the physical and dynamical parameters of Arrokoth as
obtained during the New Horizons flyby
\citep{S19LPI,C19LPI,P19LPI,Stern19Sci}.

A two-mascon model for Arrokoth shape model, with the
parameters as given in \cite{Stern19Sci} and \cite{McKi20}
provides us with the following data.

\begin{itemize}

\item The ``dumbbell size'' of Arrokoth (the distance between the
centers of masses $m_1$ and $m_2$: $d = 17.2$~km and the
radii of the components $R_1 \approx 10.1$~km, $R_2 \approx
7.3$~km.

\item Masses, assuming a typical density $\rho = 0.5$~g/cm$^3$ for cometary nuclei: $m_1 =
1.01\cdot 10^{18}$~g and $m_2 = 5.45 \cdot 10^{17}$~g. Therefore, $m_1/m_2 = 1.85$
and the reduced mass of the contact binary $\mu \equiv m_2/(m_1 + m_2) = 0.35$.

\item Rotation period of Arrokoth: $P_\mathrm{rot} =
15.92$~h, therefore $\omega = 0.77$.

\end{itemize}

\noindent The initial conditions and technical parameters are as follows:

\begin{itemize}

\item the initial positions of the two masses are set along
the $x$ axis,

\item the initial position of the test particle is at the
pericenter and its initial velocity vector (calculated in the
Arrokoth--particle two-body model) is orthogonal to the $x$ axis,

\item the maximum computation time
$T_\mathrm{max}=\omega\times10^{5}$, in Arrokoth's rotation
periods, is set in computations of the stability diagrams; and
$T_\mathrm{max}=10^{5}$, in Arrokoth's rotation periods, is set in
computations of the ejection statistics.

\end{itemize}

\begin{figure}[!t]
\centering
\includegraphics[width=0.55\textwidth]{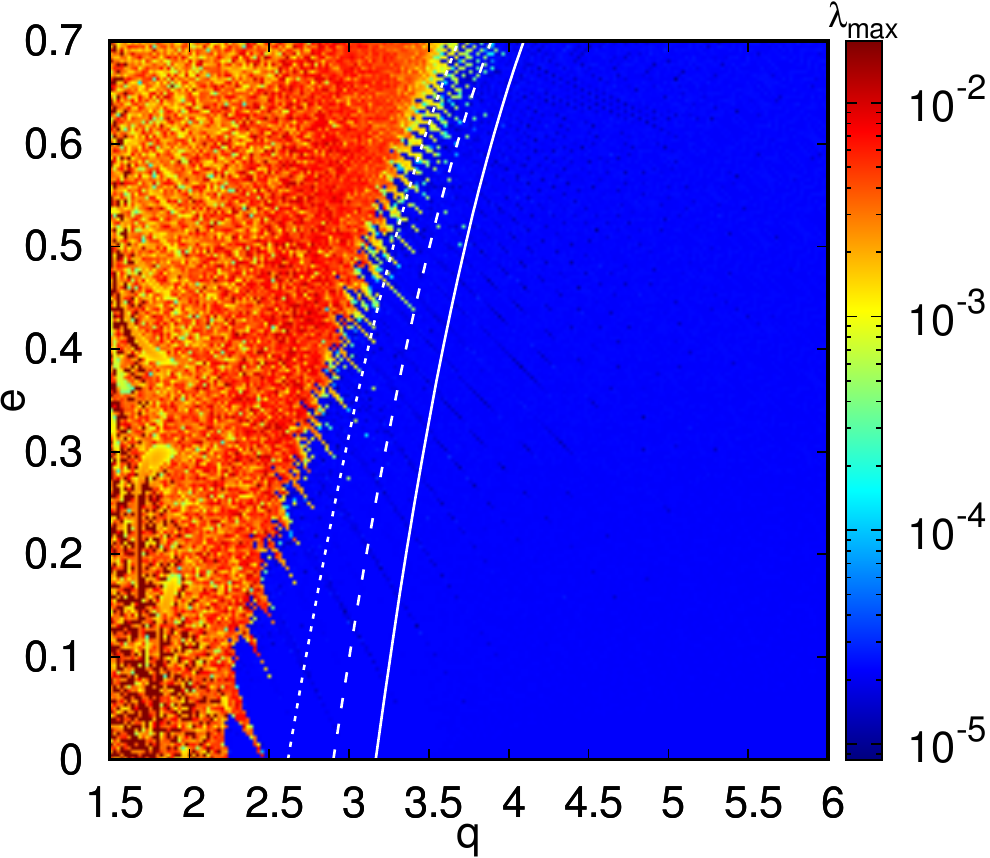}
\caption{The LCE stability diagram of the immediate dynamical
environments of Arrokoth, in finite-time LCE colour gradation.}
\label{Fig2}
\end{figure}

To build the stability diagram, 200$\times$200 orbits were computed using the Dormand--Prince
integrator DOP853 \citep{HNW87}. The local error tolerance of the integrator was set to $10^{-10}$.
The code makes a loop over $N_e$ initial values of eccentricity $e$ for any fixed initial
pericentric distance $q$. A {\it Python} code generates $N_q=200$ executables with various values
of $q$. Thus, their total number is 200.

The constructed LCE diagram of the global dynamics immediately
around Arrokoth is shown in Fig.~\ref{Fig2}. The most prominent
feature of this diagram is the ``ragged'' border between the
circumbinary chaotic zone and the outer region of regular motion.
The border is formed by the overlap of spin-orbit resonances
between the rotating Arrokoth and an orbiting particle. The most
prominent ``teeth'' of instability visible in Fig.~\ref{Fig2}
correspond to integer ratios of Arrokoth's rotation rate and an
orbiting particle's mean motion, i.e., to the $p/1$ spin-orbit
resonances.

Let $K$ be the stochasticity parameter, characterizing the overlap
of the integer spin-orbit resonances locally in the phase space of
motion, as defined in \cite{LSR18}. In Fig.~\ref{Fig2}, the solid
white curves are the theoretical borders (taking place at the
critical value $K_\mathrm{G}=0.971635406$) between the chaotic and
regular zones; the dashed curves are for $K=2$; and the
short-dashed curves are for $K=4$. These theoretical borders are
given by Equations~(6) and (11) in \cite{LSR18}. One may see that
the numerically revealed borders of chaos generally agree
with the analytical predictions: indeed, the $K=4$ analytical
curve serves approximately as a borderline above which the chaos
is complete, i.e., any regular component is negligible.

\begin{figure}[!t]
\centering
\includegraphics[width=\textwidth]{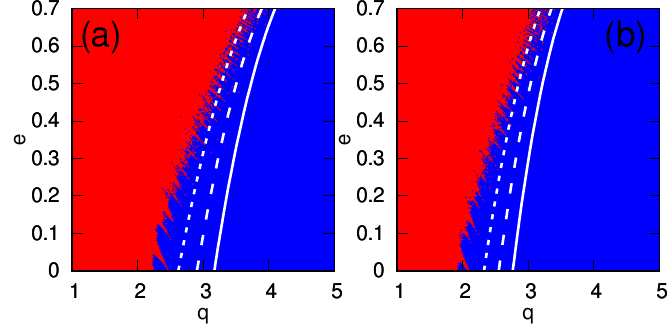}
\caption{LCE stability diagrams for the current and pre-merger
states of Arrokoth; red and blue colors correspond to chaotic
and regular orbits, respectively. Left panel (a) is for the
contact-binary phase, and right panel (b) is for a pre-merger
phase of this KBO; see text for details.} \label{Fig3}
\end{figure}

In Fig.~\ref{Fig3}, additional diagrams are constructed by means
the ``movable LCE distribution peaks'' technique. This technique
allows one to sharply separate chaotic orbits from regular ones
instead of analyzing any continuous gradation of orbits in
calculated finite-time LCE values; see \cite{SM03} for the
technique description and details.

In Fig.~\ref{Fig3}, the panel (a) corresponds to the current
(contact-binary) state of Arrokoth with the following parameters:
$\mu \simeq 0.35$ and $\omega = 0.77$. We still see
the properties described above: the ``ragged'' border and the most
prominent ``teeth'' of instability corresponding to integer ratios
of Arrokoth's rotation rate and the orbiting particle's mean
motion. For circular orbits ($e = 0$), the chaos zone size
is $\simeq 2.5$ times greater than the distance between the
two masses. The panel (b) is for a non-contact pre-merger phase;
here $\omega = 1$. Unlike the panel (a), the chaos zone size
is now $\simeq 2$ times greater than the distance between the two
masses for the circular orbits. In the both panels, the solid
white curves are the theoretical borders between chaotic and
regular zones at $K = K_\mathrm{G} \simeq 0.971635406$, the dashed
curves are built at $K = 2$, and the short-dashed curves at
$K = 4$. The theoretical borders are constructed as in
Fig.~\ref{Fig2}.

Implications of the obtained diagrams are discussed further on in
the following Sections.

\section{General background and assumptions}

Generally, the Fokker--Planck formalism can be used (adapting
approaches proposed in \citealt{MH97}, Section 3.4, and in
\citealt{T93}; see also \citealt{DQT87,MT99}) to obtain analytical
estimates of the diffusion rates in clearing processes in such or
similar systems. Here we base on the modified Kepler map
theory, as developed in \cite{LSS17,LSR18} (see also a review in
\citealt{LSS18Sch}) to describe chaotic dynamical environments of
rotating CBs.

It is important to note that our analysis is mostly developed in
the assumption that the rotation rate $\omega$ of the contact
binary is approximately the same as its critical rotation rate
$\omega_0$ of centrifugal disintegration; i.e., $\omega \sim
\omega_0$. For Arrokoth, $\omega \simeq 0.6 \omega_0$ (assuming
the typical density $\rho = 0.5$~g/cm$^3$; for smaller densities
$\omega$ would be more close to unity). This assumption allows one
to straightforwardly apply formulas already known for the case of
motion around Keplerian binaries, without any their modification.
As soon as the physical inferences made below do not mostly
require any estimates to be accurate better than by an order of
magnitude, we believe that the assumption $\omega \sim \omega_0$
is plausible for our purposes.

\begin{figure}[!t]
\centering
\includegraphics[width=\textwidth]{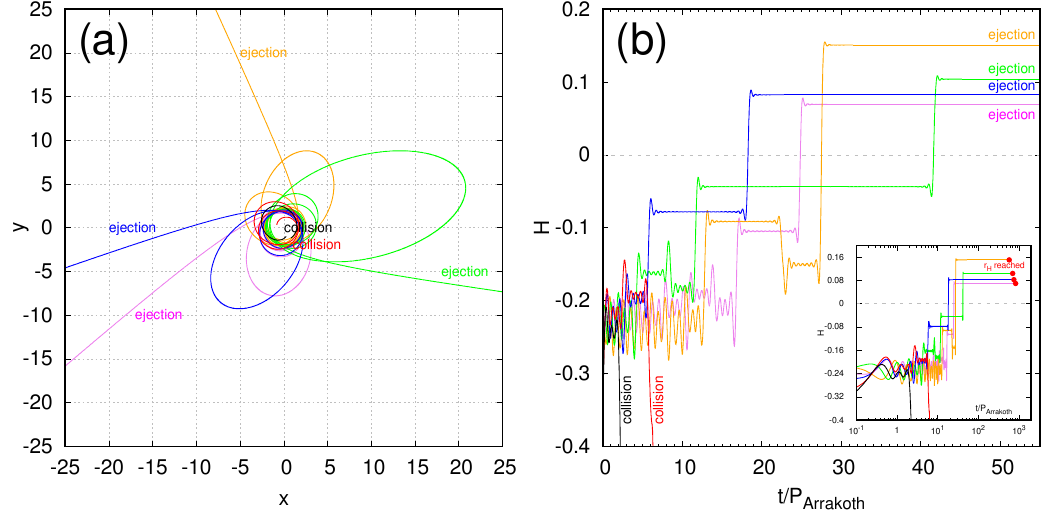}
\caption{Examples of trajectories and time evolutions of their
energies. Left panel (a): trajectories with the initial
pericentric distance $q \simeq 1.40$ (black curve),
$q \simeq 1.80$ (red), $q \simeq 2.13$ (blue),
$q \simeq 2.19$ (green), $q \simeq 2.23$ (orange)
and $q \simeq 2.27$ (purple); the initial eccentricity $e =
0$ in all cases. Right panel (b): time evolutions of the 
quantity $H = 2 | E |$ for the trajectories presented in panel
(a); the curves are coloured accordingly. The close-up: the $H$
evolutions with time in a greater range, until the particles cross
the Hill radius; the crossings are marked with red dots.}
\label{Fig4}
\end{figure}

In accord with the general scenarios of formation of contact
binaries in the Kuiper belt \citep{U19LPI,MK19LPI}, we assume
that, in the post-formation phase of Arrokoth's evolution, the
particles initially reside in a disk-like structure around the
merged binary. The theoretical circum-CB chaotic zone in
this disk may extend up to radii $\simeq 6 d$, as follows from
Figs.~\ref{Fig1} and \ref{Fig2}.

In accord with the Kepler map theory basics, we assume that, in
the motion of particles, the pericentric distance $q$ is
approximately conserved, while the semimajor axis is subject to
random walk (see \citealt{S11NA,S15ApJ}). The constancy of $q$
seems plausible down to its values of $\sim 2 d$; at smaller $q$,
the employed approximations become more and more
approximate; in particular, mergers of particles with Arrokoth
become prevalent, thus removing them. We should outline that once
$q$ (which is greater than $d$) is assumed to be constant, no
collisions with Arrokoth are possible theoretically;
therefore, the collisions are generally ignored in what follows.

To illustrate that the chaotic dynamics of particles until they
reach the Hill sphere border does indeed have a diffusive
character, in Fig.~\ref{Fig4} we present examples of trajectories
and time evolutions of their energies. In panel (a), trajectories
are shown with various initial pericentric distances $q$; the
initial eccentricity $e = 0$ in all cases. In panel (b), time
evolutions of the energies of the same trajectories are given (the
curves are coloured accordingly). The close-up shows the
quantity $H = 2 | E |$ (where the energy $E = -1/(2 a)$, and $a$
is the particle's orbital semimajor axis) evolutions with time in
a greater range, until the particles cross the Hill radius; the
crossings are marked with red dots. One may see that, for
non-collisional cases, the orbital evolution of the particles has
a random walk character (i.e., a chaotical diffusive character) in
the energy, and, therefore, in semimajor axis as well.

\section{Dispersal of matter around CBs}

For any kind of discrete motion, the diffusion coefficient $D$ can
be defined, formally, as the mean-square spread in a selected
variable (say, $H$), per time unit:

\begin{equation}
D_H \equiv \lim_{t \to \infty} \frac{\langle (H_t - H_0)^2
\rangle}{t} , \label{defD}
\end{equation}

\noindent where $t$ is time, the angular brackets denote averaging
over a set of starting values (see, e.g., \citealt{Meiss92}).

Let us define the quantity $H = 2 | E |$, where the energy $E
= -1/(2 a)$, and $a$ is the particle's orbital semimajor axis;
and the central binary's mass parameter $\mu \equiv m_2/(m_1 +
m_2)$. We extrapolate a numerical-experimental expression,
presented in \cite{DQT87} for the rate of diffusion of
circumbinary particles, from small to moderate values of $\mu$.
Taking into account that the rotation rates of the Kuiper belt
CBs, including Arrokoth, are normally of the order of the critical
rate of centrifugal disintegration (as already assumed above), one
has

\begin{equation}
D_H \simeq 100 \, H^2 \mu^2 , \label{DEdb} \end{equation}

\noindent where time is measured in pericenter passages.

For the diffusion timescale, defined as the time needed for the
particle's energy to change by an order of unity, one has

\begin{equation}
T_\mathrm{d} \simeq P \frac{H^2}{D_H} \simeq 0.01 \mu^{-2} P ,
\label{Tddb}
\end{equation}

\noindent where $P$ is the particle's orbital period averaged over
the chaotic zone.

We take $P \simeq 2 \pi a^{3/2} / ({\cal G} m_\mathrm{CB})^{3/2}$,
where $a \sim 5 d$, and $d$ is the CB's size in the mascon model.
Then, from Equation~(\ref{Tddb}) one may directly see that for a
CB like Arrokoth (with $\mu \sim 0.1$--0.3) the characteristic
timescale of the diffusion in the CB's chaotic dynamical
environment can be as small as $\sim 10$ times its rotation
period; therefore, the clearing of the chaotic zone is, in fact,
practically instantaneous.

Although our estimate of the transport time has been made in the
diffusional approximation, its smallness verifies that, 
actually, this approximation is invalid and the transport is not
diffusional, but ballistic: the clearing process is almost
``single-kick.'' This can be shown independently by calculating
the amplitude of the kick function in the Kepler map theory for
CBs, presented in \cite{LSS17,LSR18}; the kick function in the
normalized energy is given by

\begin{equation}
\Delta E\left(\mu,q,\omega,\phi\right) \simeq
W_1\left(\mu,q,\omega\right)\sin\left(\phi\right)+
W_2\left(\mu,q,\omega\right)\sin\left(2\phi\right) ,
\label{deltaEall}
\end{equation}

\noindent where $\nu = 1 - \mu$; $\phi$ is the CB's phase
when the particle is at pericenter; and the coefficients $W_1$ and
$W_2$ are given by

\begin{equation}\label{W1}
W_1\left(\mu,q,\omega\right)\simeq \mu\nu(\nu-\mu)2^{1/4}
\pi^{1/2} \omega^{5/2} q^{-1/4} \exp \left( - \frac{2^{3/2}}{3}
\omega q^{3/2} \right) ,
\end{equation}

\begin{equation}\label{W2}
W_2\left(\mu,q,\omega\right)\simeq-\mu\nu2^{15/4} \pi^{1/2}
\omega^{5/2}q^{3/4}\exp \left( -
 \frac{2^{5/2}}{3} \omega q^{3/2} \right) ,
\end{equation}

\noindent where $\omega$ is measured in units of critical $\omega_0$.

One may see that, at $\mu \sim 1/3$, $\omega \sim 1$, and $q
\sim$ 2--3, the coefficients $W_1$ and/or $W_2$ are of order unity.
The normalized single-kick energy variation is $\sim 1$, and,
therefore, indeed, an orbiting chaotic particle can be ejected in
a few kicks.

\begin{figure}[!t]
\centering
\includegraphics[width=0.85\textwidth]{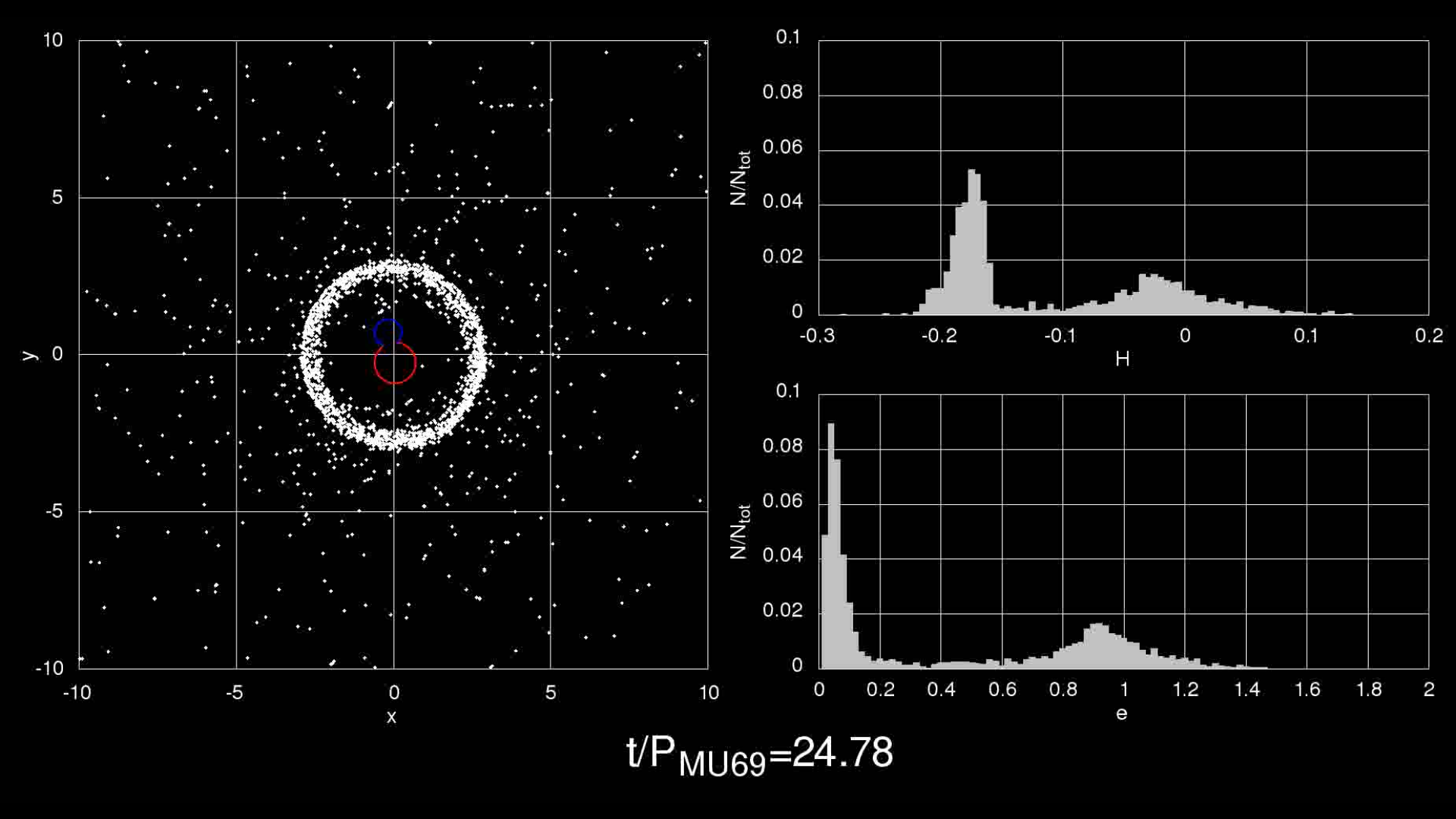}
\caption{Simulation (video) of the depopulation process of a swarm
of 10000 particles which are initially distributed in circular
orbits inside a ring $\left[1d,3d\right]$ around Arrokoth (in
the post-merger phase, for the parameters obtained
according to Fig.~1 in \cite{C19LPI}; here $\mu = 0.28$ and
$\omega = 0.59$); $H$ is measured in the barycentric
reference frame. The video can also be found at
\url{http://perso.utinam.cnrs.fr/~lages/datasets/MU69/MU69.mp4}.}
\label{Fig5}
\end{figure}

\begin{figure}[!h]
\centering
\includegraphics[width=0.75\textwidth]{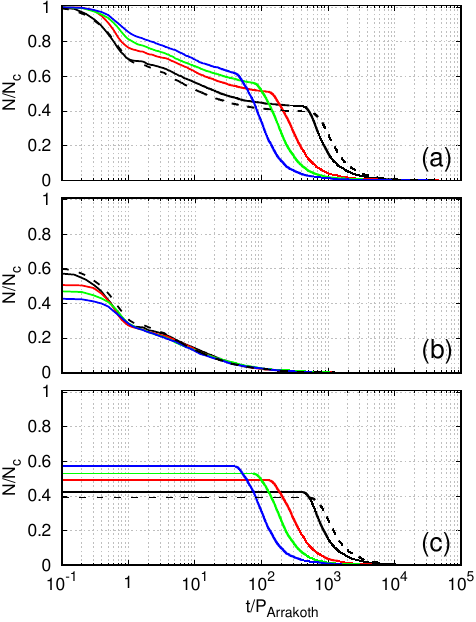}
\caption{The number of particles ejected (or colliding with
the CB) in dependence on time (counted in number of CB's
rotations). Here, $N_c$ is the number of particles initially
present in the chaotic area (particles which can exit, see
Fig.~\ref{Fig2} and Fig.~\ref{Fig3}). For each panel, the
black dashed line shows results for the Arrokoth parameters
obtained according to Fig.~1 in \cite{C19LPI}.} \label{Fig6}
\end{figure}

In Figs.~\ref{Fig5} and \ref{Fig6}, the depopulation process is
illustrated in detail, featuring several pre-merger phases of
Arrokoth. In Fig.~\ref{Fig6}, the time dependences of the number
of particles that are ejected (or collide with the CB) are
shown. The time is counted in CB's rotations. In these
simulations, 10000 particles in initially circular orbits ($e=0$)
are initially uniformly distributed in a ring with $q \in [1d,
3d]$. In Fig.~\ref{Fig6}, the black curves are for the
post-merger phase, where $d = d_0 = 17.2$~km, the radius
of $m_1$ is $r_1 \simeq 10.1$~km $\simeq 0.59d$ (at which
the collisions are fixed), the radius of $m_2$ is $r_2
\simeq 7.3$~km $\simeq0.42d$ (at which collisions are fixed), the
Hill radius is $R_\mathrm{H} \simeq 48740$~km $\simeq 3027d$ (at
which the ejections are fixed). The observational data are
taken as given in \cite{Stern19Sci,McKi20}. The rotation period
of Arrokoth is $P = 15.92$~h; this gives
$\omega=0.77\omega_{2b}$, where $\omega_{2b}$ is for the
Keplerian motion. For the post-merger phase, the complete
process of depopulation is visualized in the video provided
in Fig.~\ref{Fig5}. Note that on timescales $t$ greater than
$\sim 100P_{\rm MU69}$ the remaining particles are those initially
trapped inside the stability islet located around
$\left(q=2.5,e=0\right)$ in the phase space (the blue islet in
Fig.~\ref{Fig2}). The red curves in Fig.~\ref{Fig6} correspond to
a pre-merger phase with $d = 3d_0 = 51.6$~km, $r_1 \simeq
10.1$~km $\simeq 0.2d$, $r_2 \simeq 7.3$~km $\simeq 0.14d$,
$R_\mathrm{H} \simeq 821d$, $P = 63.5$~h; $\omega = \omega_{2b}$.
The green curves correspond to a different pre-merger phase with
$d = 5d_0 = 86$~km, $r_1 \simeq 10.1$~km $\simeq 0.12d$, $r_2
\simeq 7.3$~km $\simeq 0.08d$, $R_\mathrm{H} \simeq 492d$,
$P=136.6$~h; $\omega = \omega_{2b}$. The blue curves correspond to
a different pre-merger phase with $d = 10.1 d_0 = 172$~km, $r_1
\simeq 10.1$~km $\simeq 0.06d$, $r_2 \simeq 7.3$~km $\simeq
0.04d$, $R_\mathrm{H} \simeq 246d$, $P = 386.5$~h; $\omega =
\omega_{2b}$.

Panel Fig.~\ref{Fig6}a represents a sum of panels (b) and (c);
whereas in panel (b) the number of particles before their
collision with one of CB's components is shown. In panel (c),
statistics of particles before their ejection out from the Hill
sphere are illustrated. From Fig.~\ref{Fig6}, one may infer that
the depopulation process is rather fast already at the pre-merger
phases of the Arrokoth formation. Indeed, in Fig.~\ref{Fig6}(c),
we see that after $t \in [100,1000]$ rotations of the binary, at
least half of particles present around Arrokoth are ejected
out from the Hill sphere.

\begin{figure}[t]
\centering
\includegraphics[width=0.85\textwidth]{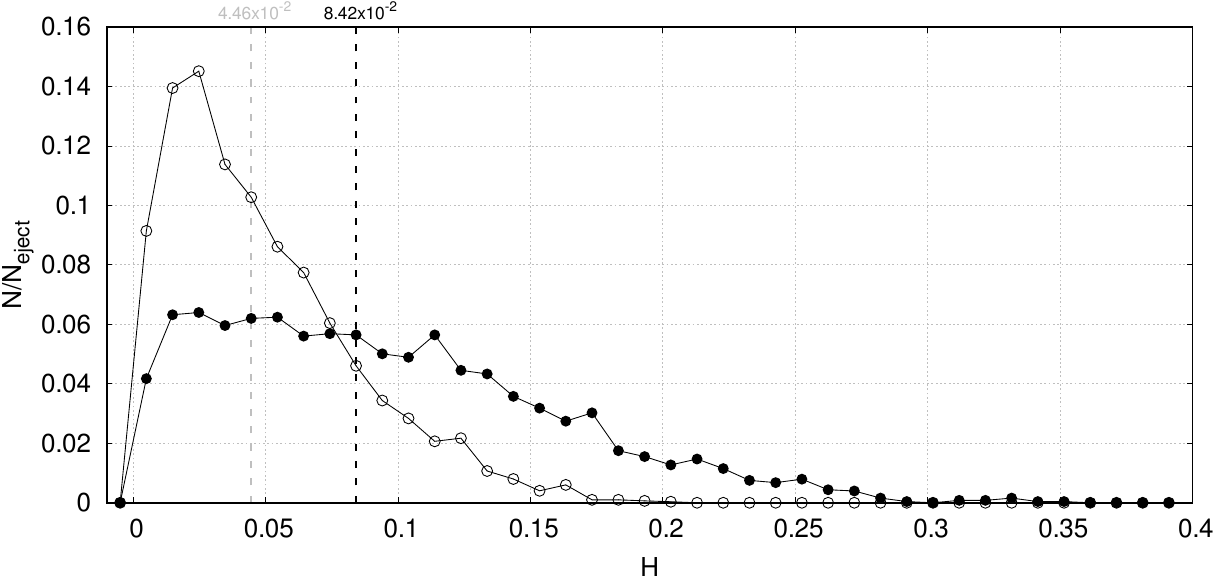}
\caption{Black dots: the distribution of $H$ for particles
crossing the Hill sphere $R_\mathrm{H}$ of Arrokoth.
Circles: results obtained using the Arrokoth parameters as
given in Fig.~1 in \cite{C19LPI}.} \label{Fig7}
\end{figure}

By inspecting the distribution of energy when the particles leave
the Hill sphere, one may calculate their ``final'' velocity value
reached upon the ejection from the Hill sphere. In
Fig.~\ref{Fig7}, the distribution of the energy $H$ on particles'
crossing Arrokoth's Hill sphere ($R_\mathrm{H} \simeq
42370$~km) is shown in the post-merger phase. Here, in our
units, where $d = 1$, is the primaries' separation and $P_{2b} =
2\pi$ is the binary's period, $H_{1/2} \simeq 0.08$, where
the first half of the ejected particles' population has $H >
H_{1/2}$ and the second half has $H < H_{1/2}$. With the previous
energy, the free particles reach $R_\mathrm{H}$ in
$R_\mathrm{H} \omega / (2\pi\sqrt{2H_{1/2}}) \simeq750$
rotations of Arrokoth. Indeed, by associating these findings with
the results given in Fig.~\ref{Fig6}, one may conclude that the
positive value of energy is reached very fast and we see that the
depopulation of the CB's disk proceeds with the typical
half-depopulation time (for the whole disk) $\sim$10--100 CB's
periods, in accord with our analytical estimate given above.

\begin{figure}[h]
\centering
\includegraphics[width=0.85\textwidth]{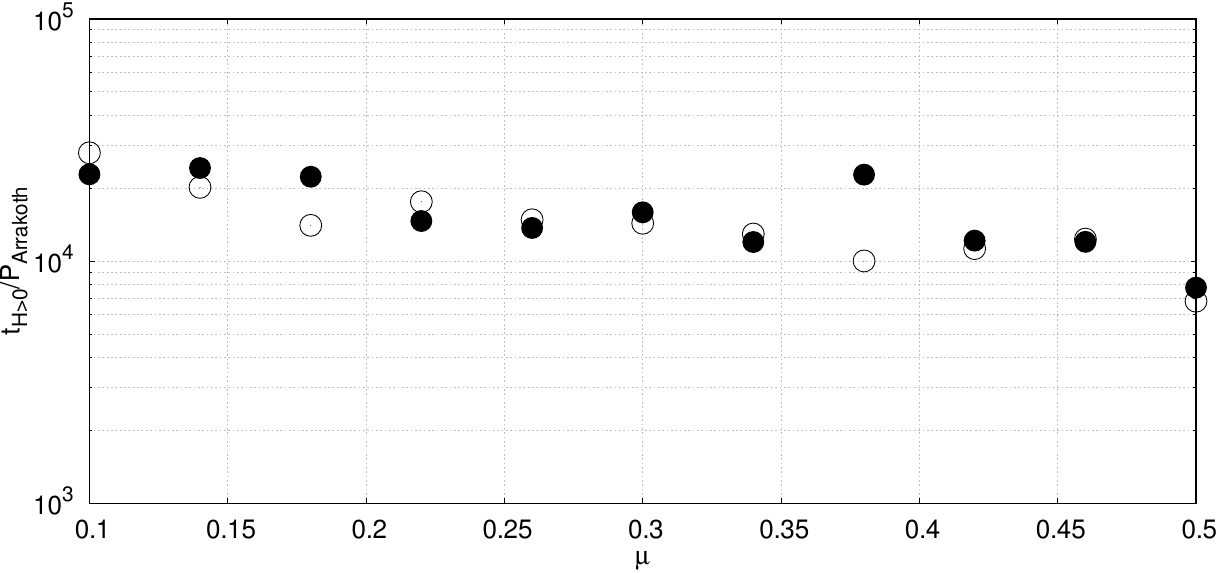}
\caption{Black dots: the mass parameter dependence for the
depopulation time $t_{H>0}$. Circles: results obtained using
the Arrokoth parameters as given in Fig.~1 in \cite{C19LPI}.}
\label{Fig8}
\end{figure}

In Fig.~\ref{Fig8}, the mass parameter dependence for the
depopulation time $t_{H>0}$ is shown. At each separate $\mu$ value
in the given range, the orbital evolution of $10^4$ particles is
simulated. For the orbits, the initial $e$ is set to zero and the
initial $q$ values are set uniformly in the interval $q \in [1d,
3d]$. Any particle is regarded as ejected, if its energy $H$
becomes positive. The depopulation time $t_{H>0}$ is fixed, when
the number of particles remaining non-ejected becomes less than
$1\%$ of the initial number of particles. One may observe that, in
the given range $0.1 \leq \mu \leq 0.5$, the depopulation time
depends on $\mu$ rather weakly, and the depopulation process is
always fast.

\section{``Mixer bowls,'' cocoons, and their long-term survival}

As we have seen above, a rotating CB is a kind of a ``cosmic
mixer,'' efficiently dispersing any neighbouring material
outwards. It is well known that any mixer (blender, eggbeater)
needs a container (a bowl) to hold the ingredients from dispersal
while mixing. Our cosmic mixer also needs such a storage bowl,
otherwise the cocoon of matter inside its Hill sphere would not
emerge.

Let us estimate the typical time $T_\mathrm{enc}$ between
encounters of relatively large KBOs (with mass or size greater
than that of Arrokoth) with Arrokoth's Hill sphere. Such
low-velocity encounters would disperse Arrokoth's cocoon, if it
were present. Therefore, if $T_\mathrm{enc}$ is much less than the
Solar system age, one can be confident that the Arrokoth's Hill
sphere is totally cleansed.

Let $N_0$ be the total number of impactors with size (radius)
greater than radius $R_0$. Taking Arrokoth's radius for $R_0$, the
characteristic (average) time between encounters of such KBOs with
Arrokoth can be written, following a general approach of
\cite{PK12ApJ}, as

\begin{equation}
T_\mathrm{enc} = (P_\mathrm{i} \sigma N_0)^{-1} , \label{Tenc}
\end{equation}

\noindent where $P_\mathrm{i}$ is the intrinsic collision
probability, measured in km$^{-2}$~yr$^{-1}$, $\sigma$ is the
collisional cross section, measured in km$^2$. For the probability
$P_\mathrm{i}$ inside the classical Kuiper Belt, there exists two
estimates: according to \cite{FDS00}, $P_\mathrm{i} = 1.3 \cdot
10^{-21}$, and, according to \cite{DMPV01}, $P_\mathrm{i} = 4
\cdot 10^{-22}$~km$^{-2}$~yr$^{-1}$.

According to Equation~(18) in \cite{PK12ApJ}, number $N_0$ of KBOs
with size $R>R_0$ can be estimated using the power-law scaling

\begin{equation}
N_0 = 618\,000 \cdot (26 / R_0)^{q-1} , \label{NR0}
\end{equation}

\noindent where radius $R_0$ is in kilometers, and we set $q
\simeq 3.5$ (a collisional equilibrium slope). With Arrokoth's
$R_0 \approx 16$~km, one has $N_0 \simeq 2.1 \cdot 10^6$.

For the collisional cross section we take the cross section of
Arrokoth's Hill sphere: $\sigma \simeq \pi R_\mathrm{H} \sim
10^{11}$~km$^2$. From Equation~(\ref{Tenc}) one finally has
$T_\mathrm{enc} \sim 4000$~yr (using $P_\mathrm{i}$ from
\citealt{FDS00}) or $T_\mathrm{enc} \sim 12\,000$~yr (using
$P_\mathrm{i}$ from \citealt{DMPV01}).

In the both cases, $T_\mathrm{enc}$ is much less than the Solar
system age. However, we have not yet taken into account that for
an encounter to disperse the cocoon, the impact velocity should be
small enough.\footnote{Paradoxically, for encounters between KBOs
themselves, an encounter is destructive if the impact velocity is,
on the opposite, high enough.} The corresponding low-velocity
threshold can be specified as the velocity at which the typical
time of traversing Arrokoth's Hill sphere by an impactor is about
the same as the typical orbital-period timescale of particles
inside the Hill sphere. If impactor's velocity is much greater
than this limit, then the cocoon remains practically unperturbed.

As derived in the previous Section, the typical orbital-period
timescale of particles inside the Hill sphere can be estimated as
$\sim 500$~yr. Given the radius of the Hill sphere $R_\mathrm{H}
\sim 5 \cdot 10^4$~km, the low-velocity limit for an impactor is
then $v_\mathrm{cr} \sim 10^{-4}$~km/s. According to \cite{FDS00}
(Table~1) or \cite{DMPV01} (Table~4), the mean impact velocity in
the Kuiper belt is about $1$~km/s. Therefore, the probability of
an impact with $v < v_\mathrm{cr}$ is much smaller than the
intrinsic impact probability. But by what amount? As follows from
\cite{DMPV01} (Fig.~7), one may assume that the frequency of
impacts at small and moderate $v$ rises with $v$ approximately
linearly (for the non-resonant population) almost up to the
maximum corresponding to the typical $v \sim 1$~km/s. Therefore,
very roughly, one may estimate that the impacts with $v <
v_\mathrm{cr}$ occur $\sim 10^4$ times less frequently than the
typical ones. As derived above, the timescale for typical impacts
is $\sim 10^4$~yr; therefore, for the dispersive low-velocity
impacts the timescale is $\sim10^8$~yr. It is still smaller than
the age of the Solar system, but not dramatically.

Counting from the Solar system early epoch, Arrokoth's cocoon
could have suffered $\sim 10$--100 dispersal events; therefore one
may expect that Arrokoth's Hill sphere (as well as Hill spheres of
similar KBOs) nowadays is empty. However, due to a number of model
approximations made, this conclusion should be subject to
verification in massive numerical simulations. In particular, it
should be taken into account that, at the Kuiper belt peripheral
regions, where the concentration of objects can be radically less,
cocoons may hypothetically survive; but this should be checked in
realistic simulations.

\section{Survival of space probes}

Let us consider in a more detail the problem of survival of space
probes visiting Arrokoth and similar objects, in the light of the
analysis performed above.

Mass $m_\mathrm{dm}$ of the debris matter left from the formation
of a given KBO is generally expected to be less than the KBO's
final mass \citep{U19LPI,MK19LPI}. Therefore, for Arrokoth,
$m_\mathrm{dm} < 2 \cdot 10^{15}$~kg. The total volume of
Arrokoth, given that $R_1 \approx 10$~km, $R_2 \approx 7$~km, is
$\sim 4(R_1^3 + R_2^3) \sim 5000$~km$^3$.

Estimating rather formally, this material, if dispersed into
boulders with size of $R_b \sim 10$~cm each, would provide $\sim
10^{15}$ boulders. When mixed inside the Hill sphere (with
$R_\mathrm{H} \sim 5 \cdot 10^4$~km), the boulder concentration
would be $\rho_b \sim 10^{-16}$~cm$^{-3}$.

In projection on a plane, this would typically result in the
column concentration $\sigma_b \sim R_\mathrm{H} \rho_b \sim
10^{-6}$~cm$^{-2}$. Given the New Horizons dimensions
$2.2\times2.1\times2.7$~m, the probe cross-section is $\sim 6
\cdot 10^4$~cm$^2$. We see that if all the dispersed material were
``conserved in the bowl,'' the probability of collisional
destruction of the space probe would be rather significant, up to
10\%. Taking smaller sizes for the dispersed boulders would raise
the probability up to unity. Since New Horizons flied away safely,
one may argue that the post-formation debris had already leaked
from the ``bowl'', or there were not much of them from the very
beginning.

Of course, this approach is strictly formal; first of all, a
realistic size distribution for fragments should be used in our
calculations. However, as soon as any realistic distribution is a
power law with the power-law index $\sim -3$, this improvement
would only produce more fragments with approximately the same
total mass; therefore, the collisional probability would only
aggravate.

\section{Conclusions}

In this article, we have explored properties of the long-term
dynamics of particles (moonlets, fragments, debris or particles)
around Arrokoth, as a prototype of many similar (dumbbell-shaped)
objects potentially present in the Kuiper belt. The chaotic
dynamics of particles inside the Hill sphere of Arrokoth (or,
generally, a similar object) has been studied numerically, by
means of construction of the LCE diagram, as well as analytically.

In the both numerical and analytical parts of our work, we have
obtained the following results.

(1)~The clearing process of the chaotic circumbinary zone is
practically instantaneous: the zone is cleared in a few ``kicks''
of the central CB.

(2)~In the studied mass parameter range $0.1 \leq \mu \leq 0.5$,
the depopulation time depends on $\mu$ rather weakly, and the
depopulation process is always fast, although it has a diffusive
character.

(3)~Due to relatively frequent low-velocity encounters of
Arrokoth's Hill sphere with other KBOs, the matter cocoon, if
formed inside the Hill sphere, could have been dispersed on a
timescale of $\sim 10^8$~yr.

(4)~If not dispersed, such a cocoon matter may pose a serious
problem for the survival of any space probe visiting Arrokoth,
since the collision probability could be well of order unity.

Our study has an implication concerning formation scenarios of
contact binaries in the Kuiper belt. As noted in \citep{U19LPI},
any such scenario, apart from producing a slowly rotating CB,
should treat how all remaining local debris are cleared away. We
underline that, irrespective of the formation scenario, the
generic chaotization of the immediate vicinities of any
gravitating ``snowman'' rotator, followed by transport processes
inside its Hill sphere, naturally explains the current absence of
such debris.

Tantalizingly, the chaotic-clearing phenomenon affects both former
targets of the New Horizons mission, but in different ways: Pluto
is not able to clean-up any radial neighbourhood of its orbit, and
on this reason it was deprived of the planetary status
\citep{IAU06}; conversely, Arrokoth is able, as we have seen
above, to create a clearing, but of another
(circum-contact-binary) kind.

This study is a first approach to the problem. The
limitations of our dynamical model include, in particular,
non-taking into account the effects due to the irregular shape of
Arrokoth's components on the orbiting particles, especially
important for particles with small pericentric distances.
(However, note that such particles, those with $q \lesssim 2 d$,
are absorbed by Arrokoth almost immediately.) Besides, the solar
gravitational effects on particles with large orbital semimajor
axes (comparable with Arrokoth's Hill radius) can be important,
although not many particles acquire large elliptic orbits around
Arrokoth; they are mostly thrown out from its Hill sphere in a
single or only several kicks (see Figs.~\ref{Fig4} and
\ref{Fig6}). The 3D study of the escape process is envisaged, and,
in the forthcoming separate study, the solar perturbations will be
taken into account.

\bigskip

\noindent {\bf \it Acknowledgments.} The authors are grateful to
Beno{\^\i}t Noyelles for helpful remarks. I.I.S. was
supported in part by the grant 13.1902.21.0039 of the Ministry of
Science and Higher Education of the Russian Federation.

\end{document}